\documentclass[reprint,prl,amsmath,amssymb,twocolumn,superscriptaddress,showpacs]{revtex4-1}
\usepackage{graphicx}
\usepackage{dcolumn}
\usepackage{verbatim}
\usepackage{bm}

\usepackage[
bookmarks=true,
colorlinks,
linkcolor=blue,
urlcolor=blue,
citecolor=blue,
plainpages=false,
pdfpagelabels,
final,
breaklinks=true
]{hyperref}

\usepackage{environ}
\NewEnviron{review}{\textcolor{red}{\BODY}}

\begin{document}

\preprint{APS/123-QED}

\title{Type-\textbf{III} corner state in second-order topological insulator by distinctly hybridized photonic Wannier functions}

\author{Zhenzhen Liu}
\author{Guochao Wei}
\author{Jun-Jun Xiao}
 \email{eiexiao@hit.edu.cn}
\affiliation{
Shenzhen Engineering Laboratory of Aerospace Detection and Imaging, College of Electronic and Information Engineering, Harbin Institute of Technology (Shenzhen), Shenzhen 518055, China
}

\date{\today}
\begin{abstract}

In the presence of crystalline symmetries, second-order topological insulators can be featured by the polarization which is believed identical to the Wannier center. In this Letter, we show that second-order topological insulators are present in the full process of topological phase transition between a pair of degenerate photonic bands resulting from the non-symmorphic glide symmetry in all-dielectric photonic crystals. Due to the coupling (hybridization) of the Wannier functions, the center of the maximally localized Wannier function always locates at the origin, not equal to the addition of constituent polarization. In this case, the corner states are ascribed to the hybridized Wannier functions obstructed by the boundary. In combination with the local density of states, the second-order topology is clearly confirmed and characterized. That ensures corner states which are fundamentally different to either the type-I corner state localized in particular sublattice, or the type-II corner state originated from interacting topological edge states. This type of corner states is regarded as type-III here and provides an alternative method to explore higher-order topological physics, which may lead to interesting applications in integrated and quantum photonics.

\end{abstract}

\maketitle

Topological physics has attracted great attentions and gained noteworthy developments \cite{Hasan2010Colloquium, Qi2011Topological, Chang2013Experimental}. Distinct topological phases have been extensively studied in various classical wave systems, e.g., photonic, phonontic and elastic systems \cite{Haldane2008Possible, Khanikaev2017Two, Ozawa2019Topological, Wu2015Scheme, Deng2017Observation, Wu2020Deterministic, Yu2018Elastic, Fan2020Elastic}. 
Beyond the traditional bulk-edge correspondence, many types of higher-order topological insulators (HOTIs) have been found which host lower-dimensional topologically protected boundary states \cite{Benalcazar2017Quantized, Benalcazar2017Electric}. For example, second-order topological insulator has gapped one-dimensional (1D) edge states and zero-dimensional (0D) corner states in the gap \cite{Xie2018Second, Zhang2019Second, Chen2019Direct}. In terms of the highly localized topological corner states, photonic nano-cavity with extremely high Q-factor \cite{Ota2019Photonic} and topological laser with low-threshold \cite{Zhang2020Low} have been designed and experimentally demonstrated.

\begin{figure}[t]
\includegraphics[width=7.5cm]{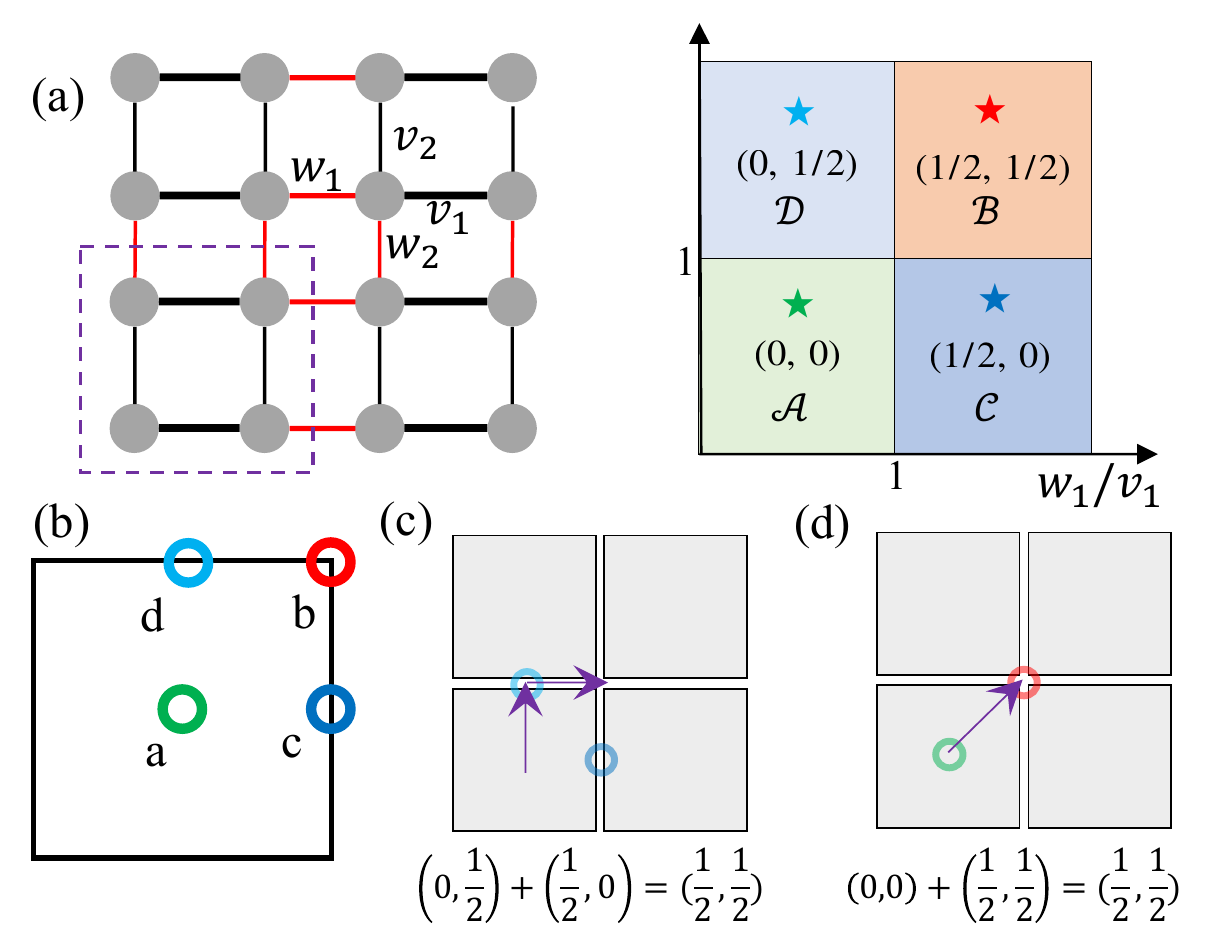}
\caption{\label{fig:fig1} (a) Schematic of 2D SSH model on a square lattice with nonsymmetric coupling in each direction and the corresponding phase diagram versus the ratio of the intra- and intercelluar coupling. The amplitudes of intra- and intercelluar coupling strength are $\emph{w}_1$ and $\emph{v}_1$ in \textit{x} direction ($\emph{w}_2$ and $\emph{v}_2$ in \textit{y} direction), respectively. (b) $C_2$ Wannier centers in (a) at Wyckoff positions (represented by hollow circles) over an individual lattice cell. The configurations of nontrivial polarization caused by stacking of models with polarization at positions (c) \textit{c} and \textit{d} and (d) \textit{a} and \textit{b}. Here, we label the configurations (c) and (d) as phase-$\alpha$ and phase-$\beta$, respectively. The equations below (c) and (d) denote the addition of polarization. The unit cell consists of four sub-sites surrounded by dashed square.}
\end{figure}

One general prototype of HOTIs is the two-dimensional (2D) Su-Schrieffer-Heeger (SSH) model~\cite{Liu2017Novel, Ezawa2018Higher, Benalcazar2019Quantization, Xie2018Second,Chen2019Direct,Ota2019Photonic}. As depicted in Fig.~\ref{fig:fig1}(a), the phase diagram of SSH model in square lattice can be approximately described by the tight-binding Hamiltonian~\cite{Liu2017Novel,Xie2018Second}. Clearly, four different phases $\mathcal{A}$-$\mathcal{D}$ as labelled in Fig.~\ref{fig:fig1}(a) and their topological phase transition can be obtained by changing the relative strength between intra- and inter-cell coupling in each direction. The topology of this tight-binding model can be characterized by the 2D dipole polarization $\mathbf{P}$ which is equivalent to the integral of the Berry connection over the entire Brillouin zone (BZ)~\cite{Liu2017Novel}. This polarization is essentially associated to the Wannier center: the center points of Wannier functions~\cite{Benalcazar2019Quantization,Peterson2020fractional}.

Figure~\ref{fig:fig1}(b) schematically shows the four elementary Wannier center configurations for the topological phases $\mathcal{A}$-$\mathcal{D}$ in Fig.~\ref{fig:fig1}(a) that satisfy $C_2$ rotation symmetry. For phase-$\mathcal{A}$ whose Wannier center is at the Wyckoff position \textit{a}, which is conventionally topologically trivial. However, Wannier centers of phase-$\mathcal{B}$, -$\mathcal{C}$ and -$\mathcal{D}$ locate between multiple unit cells (c.f., at Wyckoff position \textit{b}, \textit{c} and \textit{d}). In this regard, the fractional charge is contributed purely from Wannier centers that straddle bulk and boundary unit cell. These edge-localized fractional charges are the manifestation of first-order nontrivial topology in topological insulators~\cite{Peterson2020fractional}. Particularly, phase-$\mathcal{B}$ in Fig.~\ref{fig:fig1}(a) holds the Wannier center at Wyckoff position \textit{b} in Fig.~\ref{fig:fig1}(b), which not only contributes to fractional edge charge, but also gives rise to fractional corner charge, qualifying it as a high-order topology. The mismatch of Wannier center and lattice site produces either the type-$\bf{I}$~\cite{Xue2018Acoustic} or the type-$\bf{II}$~\cite{Li2020Higher} corner states, as widely investigated in many $C_n$ rotational systems, including $C_6$, $C_4$, and $C_3$ \cite{Li2020Higher,Benalcazar2019Quantization,Peterson2020fractional,Wu2021All,Liu2021Bulkdisclination,Fang2021Filling} and even $C_2$~\cite{Xiong2022Topological}.

However, the above-mentioned phase diagram in SSH model basically focus on one individual Wannier center-featured topological phase transition. Stacking two models from multiple bands with polarization $\mathbf{P}_1$ and $\mathbf{P}_2$ results in the phase featured by $\mathbf{P}_1+\mathbf{P}_2$~\cite{Ota2019Photonic,Zhang2019Deep}. To realize the nontrivial polarization $\mathbf{P}=(1/2,1/2)$, only two possible schemes are available, as shown in Figs.~\ref{fig:fig1}(c) and \ref{fig:fig1}(d), respectively. Notice that both of them are of the same fractional charge at edges for both \textit{x} and \textit{y} terminations (i.e., the first-order boundary). As such, both of these two cases exhibit the same bulk-edge-corner correspondence. Due to their different constituent polarization, we name these two cases as phase-$\alpha$ and phase-$\beta$, respectively. More specifically, phase-$\alpha$ hosts a corner charge $Q_c=1$, while phase-$\beta$ hosts a corner charge $Q_c=1.25$~\cite{Benalcazar2019Quantization}. Furthermore, for two degenerate bands, the coupling between two localized Wannier functions may affect the efficiency of polarization addition. In combination with the hybridized Wannier function and constituent polarization, phase-$\alpha$ and phase-$\beta$ demonstrate distinct topological property of the 0D corner states, which has not yet been clearly illustrated.

\begin{figure}[t]
\includegraphics[width=7.5cm]{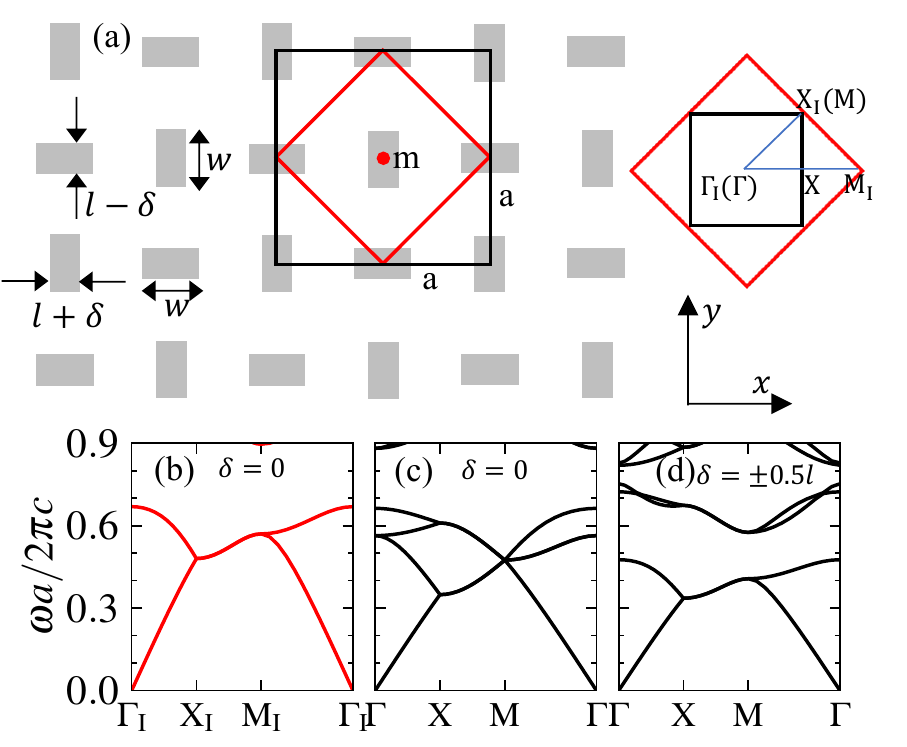}
\caption{\label{fig:fig2} Geometry and band structure of the square-lattice PCs consisting of rectangular meta-atoms with permittivity $\varepsilon_r$ and degenerate bands induced from zone folding. (a) Supercell (black box) centered at the square block (labeled as $m$). The primitive unit cell is outlined by the red box. TM band structure for (b) the primitive unit cell and (c) the enlarged supercell with $\delta=0$. (d) The TM band structure with $\delta=\pm 0.5l$. The geometrical parameters are $l=0.15a$ and $w=0.225a$. The BZ of the supercell and primitive unit cell is shown in the right panel of (a). Here $\varepsilon_r=9.5$.}
\end{figure}

To realize the proposed topological phase transition between phase-$\alpha$ and phase-$\beta$, geometric anisotropy on two-dimensional photonic crystals (PCs) is designed, as shown in Fig.~\ref{fig:fig2}(a). Based on the band folding and glide symmetry \cite{tobesubmitted,Suppl}, four-fold degeneracy as well as the topological phase transition between a pair of degeneracy can be observed. Interestingly, we find that the band gap above the lowest two bands holds the entire bulk polarization $(P_x,P_y)=(1/2,1/2)$, characterized by the parity of eigenstates at high symmetric momenta. Furthermore, the localized hybridized Wannier function is utlized to reveal the origin of topological phase-$\alpha$ and -$\beta$, and to manifest the presence of a new type of corner states, called here as type-$\mathbf{III}$.

Figure~\ref{fig:fig2}(a) shows the schematic of the 2D PCs consisting of rectangle meta-blocks whose long axis orientates alternatively toward \textit{x} and \textit{y} directions. The lattice constant is $a$, and the width and height of the rectangle blocks are, respectively, $w$ and $l$. Here, the primitive cell is outlined by the red box, whose BZ is denoted with high symmetry momentum line along $\Gamma_\mathrm{I}-\mathrm{X}_\mathrm{I}-\mathrm{M}_\mathrm{I}$. Note that the bands along the boundary $\mathrm{X}_\mathrm{I}\mathrm{M}_\mathrm{I}$ are doubly degenerated due to the combination of glide symmetry and time-reversal symmetry~\cite{Xia2019Observation}. Whereas, the unit cell outlined by the black square is enlarged, whose BZ becomes shrunken, which can be treated as the zone folding of the original BZ along the black lines~\cite{Wu2015Scheme}. Note that four-fold degeneracy at M and double degeneracy along XM are constructed, as illustrated in the transverse magnetic (TM) band structure shown in Figs.~\ref{fig:fig2}(b) and \ref{fig:fig2}(c)~\cite{Suppl}. The geometry parameter $\delta$ acting on the sizes of the blocks (e.g., $l\to l+\delta$ for $y$-oriented blocks and $l\to l-\delta$ for $x$-oriented blocks) is tuned to investigate the band evolution. When $\delta$ changes from negative to positive, the band gap closes and reopens, experiencing a topological phase transition, as shown in Fig.~\ref{fig:fig2}. These two phases are determined to be phase-$\alpha$ ($\delta<0$) and phase-$\beta$ ($\delta>0$) detailed in the following.

Due to the $C_2$ point group symmetry of our PCs, the topological properties of \textit{m}-centered PCs (denoted by $\text{PC}_{\delta}^{(m)}$) could be characterized by the 2D polarization $\mathbf{P}=(P_x,P_y)$ defined as~\cite{Fang2012Bulk, Liu2018Topological,Ota2019Photonic}
\begin{equation}\label{eq:zak}
    P_i=\frac{1}{2}(\sum_n q_i^n\  \mathrm{modulo\ 2}), (-1)^{q_i^n}=\frac{\eta_n(X_i)}{\eta_n(\Gamma)},
\end{equation}
where $i=x,y$ stands for the periodic direction and $\eta_n$ represents the eigenstate parity at the high-symmetric points $\Gamma=(0,0)$, $X_x=(\pi/a,0)$ and $X_y=(0, \pi/a)$.

The corresponding eigenstate profiles used for identifying the parities are given in the Supplemental Material~\cite{Suppl}, from which one can determine the parity by the behavior of the eigenmode profile under the inversion operation with respect to the center \textit{m} of the unit cell. In this way, the 2D polarization of the lowest two bands can be determined to be $\mathbf{P}=(1/2,1/2)$ for both $\delta<0$ and $\delta>0$. Based on the 2D polarization, the topological corner charge can be calculated as $Q_c=4P_xP_y$~\cite{Xie2018Second,Zhang2019Non,Ota2019Photonic,Zhang2019Deep}. As such, the second-order topology should be identified. To examine their difference, we focus on the realistic distributions of the hybridized Wannier functions.

\begin{figure}[t]
\includegraphics[width=8.5cm]{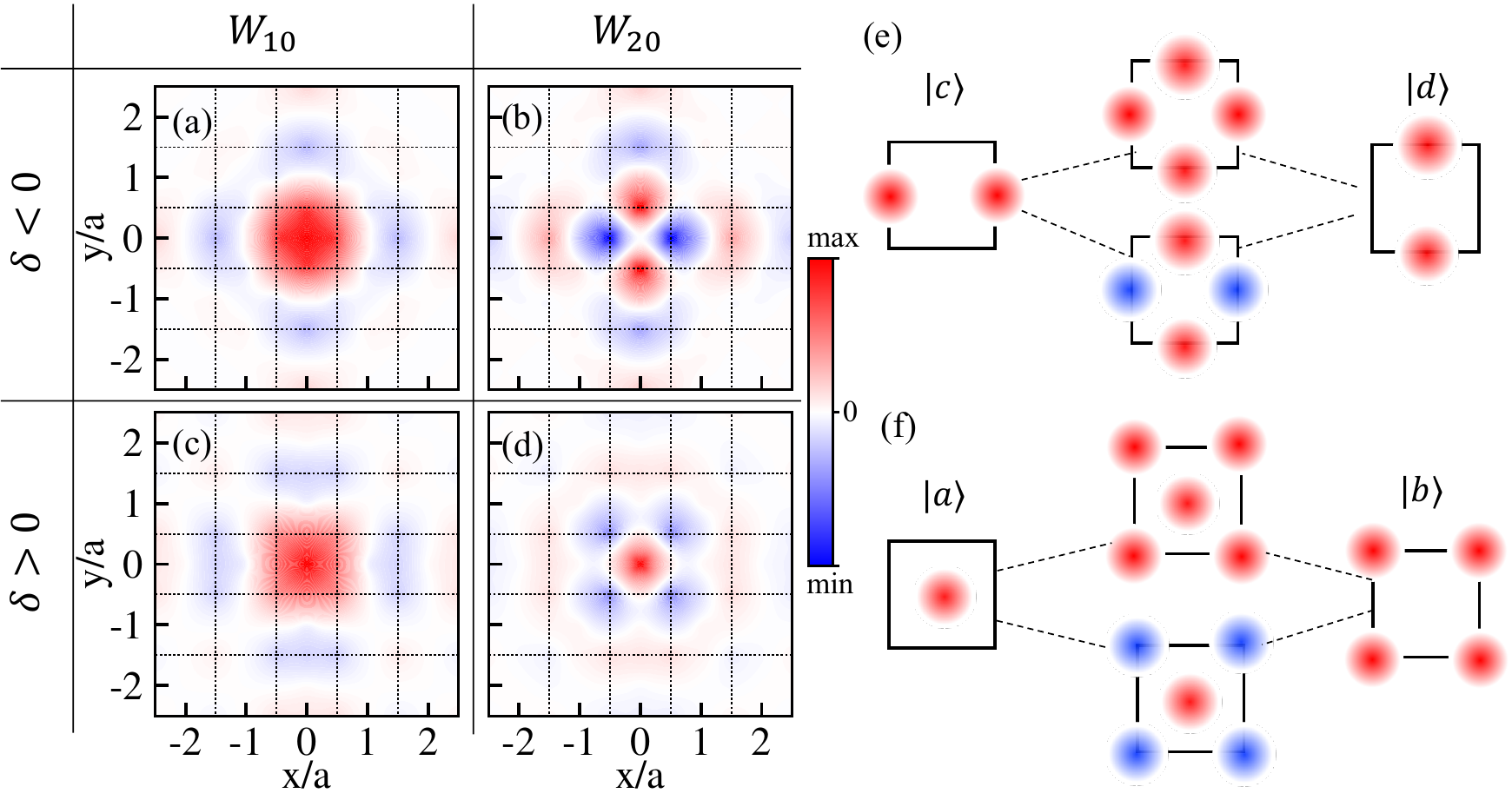}
\caption{\label{fig:fig3} MLWF $W_{n\mathbf{R}}(\mathbf{r})$ of the lowest two bands ($n=1$ or 2) for \textit{m}-centered PCs and their schematic illustration in terms of ``orbital'' coupling. (a)-(d) The hybridized Wannier functions $W_{n\mathbf{0}}(\mathbf{r})$ for $\delta<0$ and $\delta>0$. (e) Schematic illustration of two-degenerated-band (a) in-phase ($\vert c\rangle+\vert d\rangle$) and (b) out-of-phase ($\vert c\rangle-\vert d\rangle$) hybridized Wannier functions. (f) Illustration of (c) in-phase ($\vert a\rangle+\vert b\rangle$) and (d) out-of-phase ($\vert a\rangle-\vert b\rangle$) hybridized Wannier functions. Here, $+$ and $-$ indicate in-phase or out-of-phase relationship of the local ``orbitals''.}
\end{figure}

Due to the absence of $C_4$ symmetry and the presence of geometric anisotropy, the components of Wannier centers have no correlated relation. As a result, we use the maximally localized Wannier function (MLWF) to analyze topological phase transition. In 2D PCs, the MLWF is defined by the unitary transformation that minimizes the spread functional~\cite{Marzari2012Maximally}. This can be done by using the steepest descent method \cite{Marzari2012Maximally,Busch2011photonic}. Figures~\ref{fig:fig3}(a) and \ref{fig:fig3}(b) show the MLWFs for the case of $\delta<0$, and Figs.~\ref{fig:fig3}(c) and \ref{fig:fig3}(d) are for $\delta>0$. The four MLWFs are highly localized and centered at the origin. This contradicts with the nontrivial polarization obtained by Eq.~(\ref{eq:zak}). It is the band degeneracy that causes the individual local ``orbitals'' to be hybridized. More specifically, for $\delta<0$ the in-phase ($\vert c\rangle+\vert d\rangle$) and out-of-phase ($\vert c\rangle-\vert d\rangle$) coupling between the local ``orbitals'' $\vert c\rangle$ and $\vert d\rangle$ shown in Fig.~\ref{fig:fig3}(e) are consistent with the calculated MLWFs [see Fig.~\ref{fig:fig3}(a) and \ref{fig:fig3}(b)] obtained by finite element method (FEM). In contrast to the Wannier functions which are usually localized inside the structure cell, these MLWFs are mainly localized at the cell edges. Similarly, Fig.~\ref{fig:fig3}(f) illustrates the coupling result, i.e., the in-phase ($\vert a\rangle+\vert b\rangle$) and out-of-phase ($\vert a\rangle-\vert b\rangle$) hybridized ``orbitals'' for the case of $\delta>0$. The situations are also in consistent with the numerically obtained MLWFs [see Fig.~\ref{fig:fig3}(c) and \ref{fig:fig3}(d)]. In this case, the MLWFs are mainly localized at the cell corners and the center (e.g. origin). In view of the coupling configuration, all the MLWFs in Fig.~\ref{fig:fig3} are shared by adjacent cells. Therefore, the MLWFs are obstructed by the boundaries, producing the edge and corner states. Then we can conclude that instead of by simply referring to the polarization, distinctly hybridized MLWFs are responsible for the high-order topology.

Physically, the spectral charge defines the number of photonic modes in a local cell. This can be extracted by integrating the local density of states (LDOS) up to the bulk band gap \cite{Liu2021Bulkdisclination}. Due to the presence of boundaries, the Wannier functions in the edge unit cell are obstructed by the edge boundary and generate edge states. For a corner cell, the Wannier functions localized at \textit{b}, \textit{c} and \textit{d} are obstructed by the boundary and generate edge and corner states. To verify these analysis and predictions, we examine the eigenstates in a finite-sized ($10\times 10$) 2D PCs enclosed by perfect electric conductors (PECs), with a thin air gap $g$. The details of the calculation on spectral charge and LDOS are in the Supplemental Material~\cite{Suppl}.

\begin{figure}[tb]
\includegraphics[width=8.5cm]{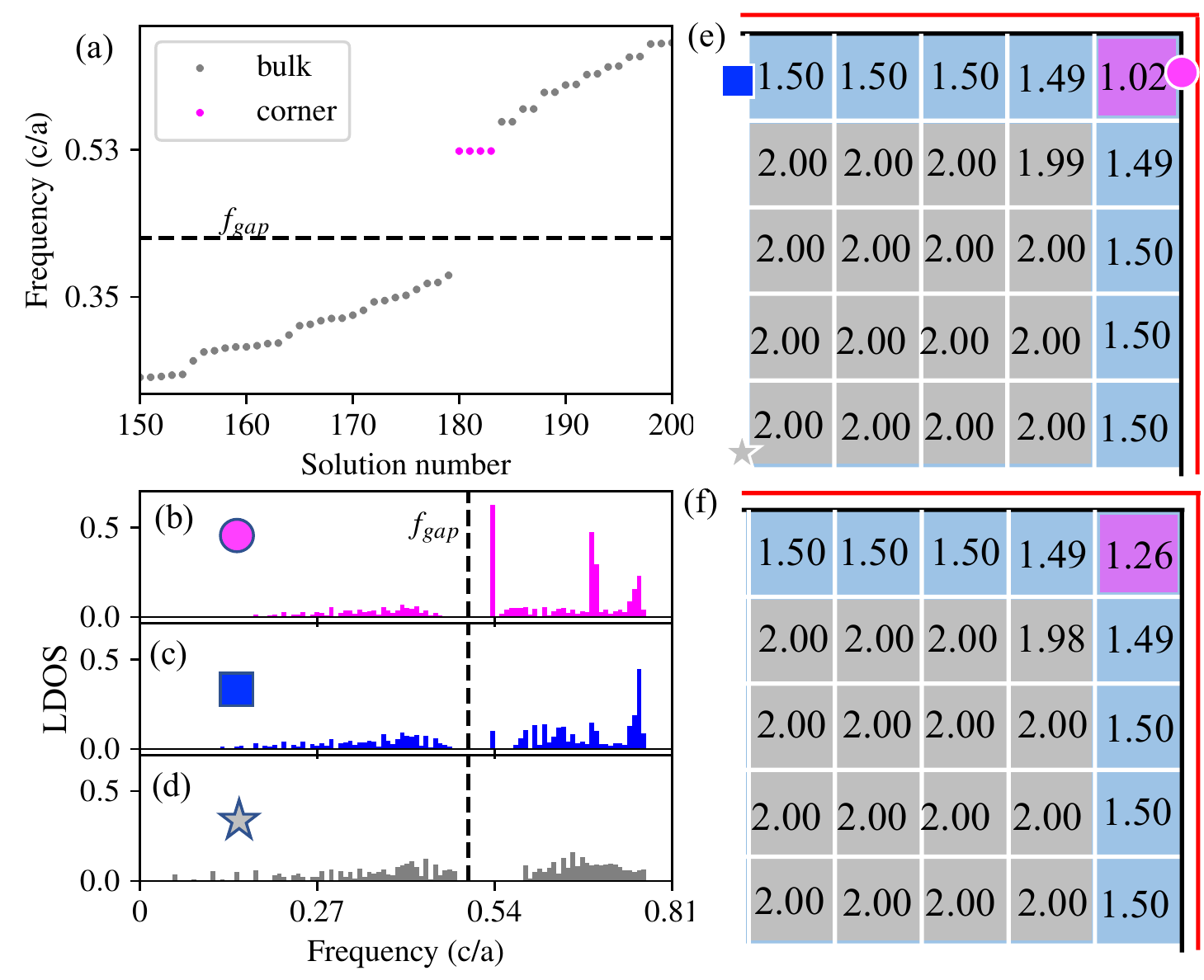}
\caption{\label{fig:fig4} (a) Photonic spectrum around the bulk band gap for finite-sized \textit{m}-centered PC with $\delta=-0.8l$ enclosed by PECs. The gray region denotes the bulk, and the magenta region denotes the corners. (b)-(d) LDOS for photons in different types of unit cells, respectively, labeled by the circle (corner unit cell), square (edge unit cell) and the star (bulk unit cell). Distribution of spectral charges in finite samples with (e) $\delta=-0.8l$ and (f) $\delta=0.8l$.}
\end{figure}

\begin{figure*}[htb]
\includegraphics[width=15cm]{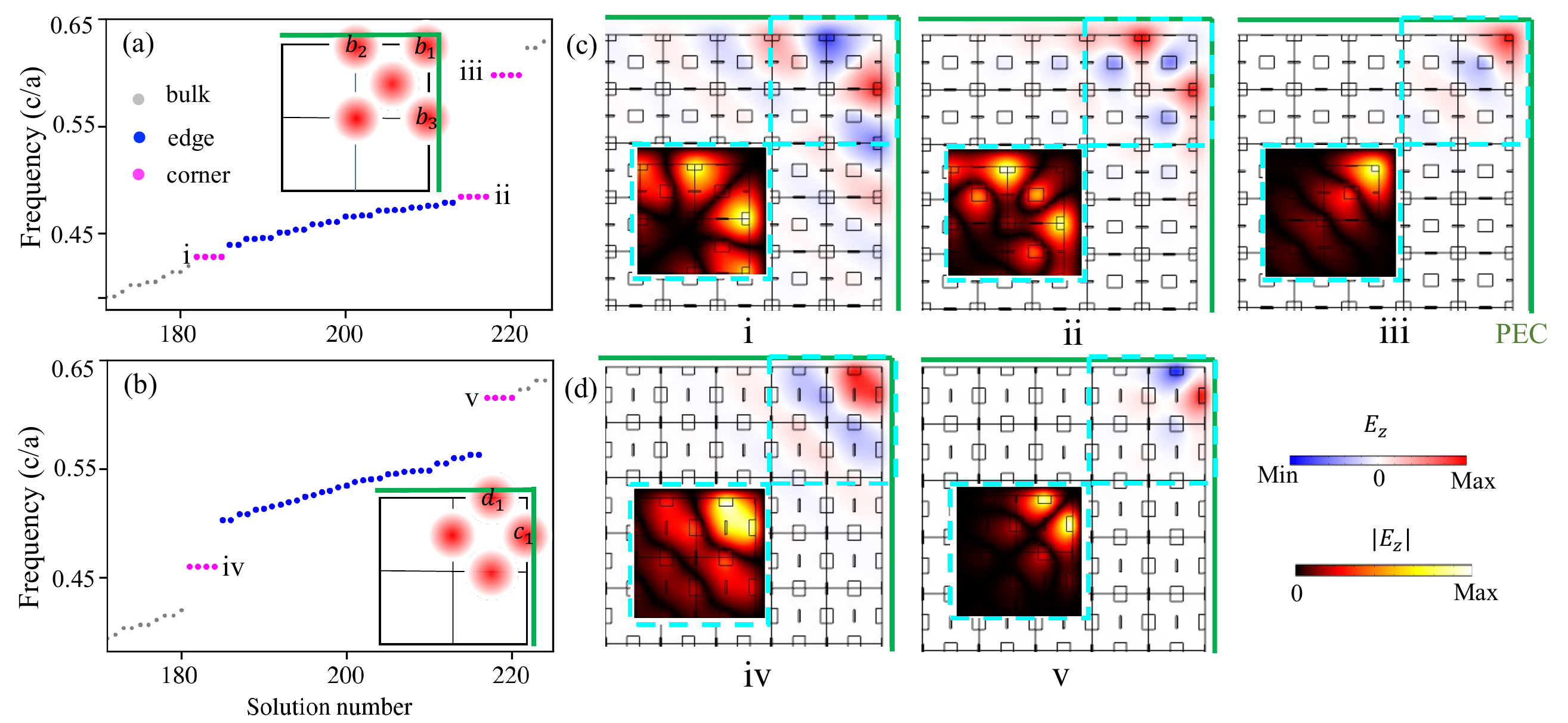}
\caption{\label{fig:fig5} Band structure of a finite sample enclosed by PEC boundaries and the field distribution of the corner states. (a) Phase-$\beta$ within $\mathrm{PC}_{0.8l}^{(m)}$. (b) Phase-$\alpha$ within $\mathrm{PC}_{-0.8l}^{(m)}$. (c) The field profiles of the corner states (\textbf{i}, \textbf{ii}, \textbf{iii}) labeled in (a), showing first, second, and zero-order characteristic, respectively. (d) The field profiles of the corner states (\textbf{iv}, \textbf{v}) labeled in (b). The insets in (a) and (b) are the corresponding suborbitals at the sample corner. The inserts in (c) and (d) are the corresponding intensity profile $\vert E_z\vert$. The distance between the PCs and PEC are (a) $g=0.3a$ and (b) $g=0.15a$, respectively.}
\end{figure*}

Note that only bulk states contribute to the spectral charge. Then from the perspective of the schematic coupling as shown in Fig.~\ref{fig:fig3}(e), the hybridized Wannier function within one bulk contributes 2 bulk states for $\text{PC}_{\delta<0}^{(m)}$. While, an edge (corner) cell contributes 1.5 (1) bulk state. This can be verified by the calculation of LDOS shown in Fig.~\ref{fig:fig4}(b)-(d). Numerically calculated ``spectral charges'' shown in Fig.~\ref{fig:fig4}(e) are close to the theoretical values. Then, the full finite sample contributes totally 180 bulk states below the frequency $f_{gap}$, as shown in Fig.~\ref{fig:fig4}(a). A similar analysis can be applied to Fig.~\ref{fig:fig3}(f) for $\text{PC}_{\delta>0}^{(m)}$. The difference is that the corner cells contribute only 1.25 bulk states, numerically verified by Fig.~\ref{fig:fig4}(f). More details are in Supplemental Material~\cite{Suppl}.

Due to their unique profile of the hybridized Wannier function for phase-$\alpha$ and -$\beta$, the topological corner states exhibit different properties. Figure~\ref{fig:fig5} shows the eigenspectrum and typical eigenstates of a finite sample with different types of corner states at the band gap. This is obtained by setting different distance between the PCs and PECs, i.e., $g=0.3a$ for $\delta=0.8l$ in Fig.~\ref{fig:fig5}(a) and $g=0.15a$ for $\delta=-0.8l$ in Fig.~\ref{fig:fig5}(b). These corner states can be transformed into the bulk when the gap \textit{g} is changed (see details in Supplemental Material \cite{Suppl}). Figures~\ref{fig:fig5}(c) and \ref{fig:fig5}(d) show the field distributions $E_z$ of different corner states, labeled in Figs.~\ref{fig:fig5}(a) and \ref{fig:fig5}(b), respectively. Here only the top-right corner localized states are shown to exemplify the topological properties of these states. Clearly, these corner states exhibit distinguishable profiles and different localization.

Here, the obstruction of hybridized Wannier functions is the cause to the presence of corner state and the distinct type of corner states. As shown in Fig.~\ref{fig:fig5}(c), corner states, -${\bf i}$, -$\bf{ii}$ and -$\bf{iii}$, emerge at the band gap, exhibiting first, second and zero-order characteristics, respectively. Particularly, the obstruction of the Wannier function localized at suborbital $b_1$ as shown in the inset of Fig.~\ref{fig:fig5}(a) results in a corner state localized at the corner of the corner cell [see Fig.~\ref{fig:fig5}(c) panel $\bf{iii}$]. This is the typical type-\textbf{I} corner state. Simultaneously, the obstruction of the suborbitals $b_2$ and $b_3$ and their coupling contributes to the corner states with odd and even parity [see Fig.~\ref{fig:fig5}(c) panels $\bf{i}$ and $\bf{ii}$]. Since $b_2$ and $b_3$ are both within the unit cell at the structure corner and edge, the resultant corner states are partially embedded into the edge modes. The coupling effect yields the so-called type-$\bf{II}$ corner states, which have been generally understood as the long-range coupling between different edges~\cite{Li2020Higher,Xiong2022Topological}. While for phase-$\alpha$ case as shown in Fig.~\ref{fig:fig5}(b) and \ref{fig:fig5}(d), only two typical corner states, shown in panel-${\bf iv}$ and -$\bf{v}$, exist. Clearly, these two corner states are mainly localized in the corner cell and they are not coupled to any edge state. In terms of the hybridization configuration of the Wannier function shown in the inset of Fig.~\ref{fig:fig5}(b), the presence of the PEC boundary leads to the obstruction of suborbitals $c_1$ and $d_1$. Here, both of the two suborbitals belong to the structure's corner cell and they have little effect on the adjacent edge and bulk cells. In this way, the resultant corner state looks visually like a quarter of the hybridized Wannier function [see Fig.~\ref{fig:fig3}(a) and \ref{fig:fig3}(b)]. Since they are different to the type-\textbf{I} and -\textbf{II} corner states, here, we regard it as type-\textbf{III}. Due to their distinct origin rooted in the hybridized Wannier functions, these corner states exhibit different localization strengths, in decreasing order $\bf{iii}-\bf{v}-\bf{iv}-\bf{ii}-\bf{i}$. This is numerically shown by the inverse participation ratio as detailed in Supplemental Material~\cite{Suppl}.

In summary, we demonstrate a kind of second-order topological insulators with two distinct Wannier centers in geometrically anisotropic 2D PCs. Hybridized Wannier functions of two degenerate bands are used to distinguish and reveal rich topological phases and their transition. Three types of 0D localized corner states with different characteristics are present in such system. Furthermore, we show the presence of one type of corner states that stems form the unique Wannier functions, instead of by the long-range edge mode interactions. Our work suggests that Wannier function in highly geometric-anisotropic system may generate interesting topological physics beyond existing description. It is expected that more exotic Wannier function might promise distinctly topological protected state and more interesting topological photonic phenomena. For instance, it might be possible to involve gauge field to specific lattice directly with magnetoelectric response by gyromagnetic material or artificially.
\newline

\begin{acknowledgments}
This work was financially supported by Shenzhen Science and Technology Program (No.~JCYJ20210324132416040), Guangdong Natural Science Foundation (No.~2022A1515011488), and National Key R$\&$D Program of China (No.~2018YFB1305500).
\end{acknowledgments}

%

\end{document}